\documentstyle[prb,aps,twocolumn,epsf]{revtex}
\begin{document}
\twocolumn[\hsize\textwidth\columnwidth\hsize\csname@twocolumnfalse%
\endcsname
\title{Non-equilibrium DC noise in a Luttinger liquid with impurity}
\author{P. Fendley and H.
Saleur}
\address{Department of Physics, University of Southern
California,
Los Angeles CA 90089-0484}
\date{January 1996, cond-mat/9601117}
\maketitle
\begin{abstract}

We compute exactly the non-equilibrium DC noise in a Luttinger liquid
with an impurity and an applied voltage. By generalizing Landauer
transport theory for Fermi liquids to interacting, integrable systems,
we relate this noise to the density fluctuations of quasiparticles. We
then show how to compute these fluctuations using the Bethe
ansatz. The non-trivial density correlations from the interactions
result in a substantial part of the non-equilibrium noise.  The final
result for the noise is a scaling function of the voltage, temperature
and impurity coupling.  It may eventually be observable in tunneling
between edges of a fractional quantum Hall effect device.

\end{abstract}
\pacs{PACS numbers:  ???}
]

\section{Introduction}

The study of fluctuations in systems of non-interacting particles is a
well-understood problem.  Of particular interest is the noise due to
free electrons scattering off an impurity, which has been discussed
carefully in \cite{all,Land,Butt}.  In these calculations, the only
interactions are those by individual particles with a barrier. The
fact that the particles do not interact with each other is of course
crucial. When interactions are present and the system is out of
equilibrium, it is not possible in general to go beyond a perturbative
analysis of the fluctuations. In particular, the issue of possible
singularities in the noise is very difficult to resolve \cite{CFW}.

In this paper, we present an exact calculation of the non-equilibrium
noise in a one-dimensional (interacting) Luttinger liquid with a
single impurity.  This problem has attracted a great deal of attention
because the edge of a fractional quantum Hall device should be a
Luttinger liquid \cite{Wen,Moon,KF}. A point contact pinching the
sample plays the role of the impurity, with the gate voltage allowing
the impurity coupling to be varied.  The noise here is especially
important because DC transport measurements are not sensitive to the
charge of the carriers, while non-equilibrium noise measurements do
depend on this charge \cite{KFnoise}. We obtain the general current
(and voltage) noise as a function of the applied voltage, temperature
and impurity coupling. We check in particular that the current noise
corresponds to tunneling of uncorrelated electrons in the
weak-backscattering limit, and to tunneling of uncorrelated Laughlin
quasiparticles of fractional charge in the strong-backscattering
limit.

In a critical theory, conformal invariance usually yields the
correlators, from which the fluctuations follow.
The situation changes dramatically when an impurity is introduced.
This adds interactions to the model and destroys conformal
invariance.
One can use the Keldysh formalism to relate various types of
fluctuations to each other, and to obtain results to lowest order in
perturbation theory in the interaction strength
\cite{KFnoise}.
Perturbed conformal field theory
yields other information \cite{Wennoise,CFW}\ (in particular it
predicts
a singularity at finite frequency), but in neither approach
has it been possible to compute exactly the non-equilibrium  noise
in the presence of interactions.

In a series of papers with A.\ Ludwig,
we have developed a description of a Luttinger liquid
in terms of its interacting quasiparticles.
As in our earlier work, the results rely
crucially on the fact that the Luttinger model with impurity is
integrable \cite{GZ,FLS}.
This formalism makes it possible
to do exact non-perturbative calculations in the presence of the
impurity.
We have computed the current and conductance through the impurity
\cite{FLS,FLSbig}, and the DC noise at zero temperature
\cite{FLSnoise}.
At zero temperature, the fluctuations are pure shot noise coming from
the impurity, so finding them does not require understanding the
non-trivial quantum fluctuations of the interacting quasiparticles.
However, at non-zero temperature, the thermal noise
and the shot noise can not be separated,
so the full noise requires understanding these fluctuations.
The effect of these fluctuations
is substantial; without including them one does not even
obtain the equilibrium
Johnson-Nyquist formula when the impurity is not
present.
In this paper, we calculate exactly the
quasiparticle fluctuations, and use them to find the exact DC noise
at
any temperature, voltage and impurity coupling. 

In sect.\ 2, we show how to compute density fluctuations
of the quasiparticles in an integrable theory, and apply the results
to find the current noise in
the Luttinger model without an impurity. In sect.\ 3,
we couple an impurity to the Luttinger liquid, and find the exact
scaling
curve. This involves  a reformulation of Landauer's approach suitable
for interacting quasiparticles.  In sect.\ 4, we consider various
limits, including the weak and strong backscattering limit, and
also derive some generalized fluctuation-dissipation results.

\section{Density and current fluctuations in an integrable theory}

In this section we first review known results for the
quasiparticle densities in an integrable theory. We
extend these calculations to compute the density fluctuations.
We then specialize to the case of interest, the Luttinger liquid
(in this section, without an impurity).
For this model, we find the current and charge fluctuations.

\subsection{General considerations}

We consider an integrable theory whose space of states is described
entirely in terms of quasiparticles.
In an integrable theory, the scattering is
completely elastic and factorizable, which means that the energy
and momentum of the quasiparticles are ${\it individually}$
conserved,
and that the multi-particle $S$ matrix decomposes into a product
of two-particle elements. For simplicity we require that
the scattering is diagonal as well, which means that
the types of the particles do not change in the collision. This
does not mean the scattering is trivial, because
there can be a phase shift; the $S$ matrix element
$S^{bulk}_{ij}(p_1,p_2)$ for scattering a particle of type $i$ with
one of type $j$ is a non-trivial function of the momenta.
It will be convenient to parametrize the momenta in terms of
the rapidity $\theta$ defined for gapless theories as $p\propto
\exp\theta$. Then scale invariance
requires that $S$ matrix element is a function of $p_1/p_2$,
or equivalently, $\theta_1-\theta_2$. (In a massive theory one
has $p\propto \sinh\theta$ and relativistic invariance requires
also that $S$ is a function of $\theta_1-\theta_2$; the results
of the section apply to this case as well.)

We study the system
on a circle of large length $L$ and at temperature $T$.
Each quasiparticle has the energy $e_i$; in a gapless theory
$e_i=\pm p_i \propto \exp\theta$.
We introduce the level densities $n_i(\theta)$ and the filling
fractions
$f_i(\theta)$, defined so that $L n_i(\theta)d\theta $ is the number
of allowed states for a quasiparticle of type $i$
in the rapidity range between $\theta$ and $\theta+d\theta$
and $f_i(\theta)$ is the fraction of these which are occupied.
Thus the density of occupied states per unit length $P_i(\theta)$
is given by
$$P_i(\theta)=n_i(\theta)f_i(\theta).$$
Requiring that the wave functions be periodic
in space gives rise
to the Bethe equations (setting $h=1$, not the usual $\hbar=1$)
\begin{equation}
n_i(\theta)={dp_i\over d\theta}+\sum_j
\int d\theta'\ \Phi_{ij}(\theta-\theta') P_j(\theta'),
\label{bethe}
\end{equation}
where the kernel is
defined by
$$\Phi_{ij}(\theta)={1\over 2i\pi}{\partial\over \partial \theta}\ln
S^{bulk}_{ij}(\theta),$$
and all rapidity integrals in this paper run from $-\infty$ to
$\infty$.
These Bethe equations hold for any allowed configuration, whether it
is
the equilibrium configuration or not. Notice that
in a free theory the kernel vanishes and we obtain the
free-particle
relation $ n_i(\theta)={dp_i/d\theta}$.

We generalize the usual analysis by allowing for
a chemical potential which depends on $\theta$, thus
contributing
\begin{equation}
\nonumber
\exp\left[L\sum_i \int d\theta \mu_i(\theta) P_i(\theta)\right]
\end{equation}
to the Boltzmann weight of a configuration.
This generalization
may seem a bit odd, but in an integrable theory with diagonal
scattering, collisions change neither
the energy nor type of a particle. Therefore $P_i(\theta)$
is independent of time, even when a current is flowing. In the
next section we discuss how the densities change with time, but
in the very special situation where the change results
from a single impurity.

To determine the equilibrium values of the densities, we
find the configuration which minimizes the free energy.
This procedure is known as the thermodynamic Bethe ansatz \cite{YY},
and is discussed in the  appendix.
We define the ``pseudoenergies'' $\epsilon_i$ by
\begin{equation}
f_i={P_i\over n_i}\equiv
 {1\over 1+e^{\epsilon_i-\mu_i}},
\label{deff}
\end{equation}
where $\epsilon_i$ and the chemical potentials $\mu_i$
are scaled by the temperature to make them dimensionless.
The result is the thermodynamic Bethe ansatz
equations
\begin{eqnarray}
\nonumber
\overline{\epsilon}_j(\theta)
&=&-\sum_k \int d\theta'\
\Phi_{jk}(\theta-\theta')
\ln\left[1+e^{-\overline{\epsilon}_k(\theta')+
\mu_k(\theta')}\right],\\
&&\qquad +{e_j(\theta)\over T}
\label{tbaeqs}
\end{eqnarray}
The bar indicates thermal equilibrium values.
The equilibrium value $\overline{F}$ of the free energy per unit
length
is
\begin{equation}
\overline{F} = -T\sum_i \int d\theta\ e_i(\theta) \ln\left[
1+ e^{-\overline{\epsilon}_i(\theta)+\mu_i(\theta)}\right].
\label{feq}
\end{equation}

The relations (\ref{tbaeqs}) and (\ref{feq})  still hold
when the $\mu_i$ depends on $\theta$.
This makes it possible to compute correlators by taking
functional derivatives of the free energy
with respect to $\mu_i(\theta)$, and
then setting $\mu_i$ to the appropriate value.
This is because the partition function can be written as
the functional integral
$$Z=\int \left[\prod_i {\cal D}P_i(\theta)\right]
 e^{-F_0L/T+L\sum_i \int d\theta \mu_i(\theta) P_i(\theta)},$$
where the explicit form of $F_0$ is given in the appendix  (here,
the exact form of the integration measure is somewhat unclear, but
this
will not affect the following analysis).
The equilibrium value $-\overline{F}L/T$ is the saddle-point value of
the
exponent. Taking derivatives with respect to $\mu_i(\theta)$ brings
down powers of $P_i(\theta)$ and gives expectation values.
For example, the equilibrium particle densities are given by
\begin{equation}
\overline{P}_i(\theta) = {1\over ZL}
{\delta Z
\over \delta \mu_i(\theta)}=-{1\over T} {\delta \overline{F}
\over \delta \mu_i(\theta)}.
\label{onept}
\end{equation}
Of course doing this functional derivation is not necessary for
determining $\overline{P}_i$; a direct way is to relate it to
$\overline\epsilon$
using the definition (\ref{deff}) and the Bethe equations
(\ref{bethe}).

To find the current noise we will need the density-density
correlation
function, which is given by second-order functional derivatives:
$$
\overline{ P_i(\theta) P_j(\theta')} =
{1\over ZL^2} {\delta^2 Z
\over \delta \mu_i(\theta)\delta\mu_j(\theta')}
$$
It is convenient to define the difference of an quantity and its
equilibrium value, e.g.\ $\Delta P_i \equiv P_i  - \overline{P}_i$.
We then have
\begin{equation}
\overline{\Delta P_i(\theta) \Delta P_j(\theta')} =
-{1\over LT} {\delta^2 \overline{F}
\over \delta \mu_i(\theta)\delta\mu_j(\theta')}
\label{dd}
\end{equation}
In general, this equilibrium value is not diagonal: it can be
non-zero
even when $\theta\ne\theta'$ and $i\ne j$. 
The diagonal correlations come from the Fermi repulsion
of the particles, while the non-diagonal part arises from
the interactions. However, we show in the appendix
that in integrable systems there is at least one kind of diagonal
fluctuation, even if interacting:
$$
\overline{\Delta f_i(\theta)\Delta f_j(\theta')} \propto \delta_{ij}
\delta(\theta-\theta')
$$It would be interesting to find a deep reason for this relation.

There is a simple alternative formulation
of (\ref{dd}) in terms of the $\overline{\epsilon}_i$. We discuss
this
in the appendix . The result is written in terms of a function
$K_{ij}$
defined by
\begin{equation}
K_{ij}(\theta,\theta')\equiv \Phi_{ij}(\theta-\theta')-
(1+e^{\overline{\epsilon}_i(\theta)-\mu_i(\theta)})
\delta_{ij}\delta(\theta-\theta').
\label{Kdef}
\end{equation}
Then we find that
\begin{equation}
\overline{\Delta P_i(\theta) \Delta P_j(\theta')}={T\over L}\left(
{\partial
\over \partial \theta} + {\partial
\over \partial \theta'}\right) K^{-1}_{ij}(\theta,\theta'),
\label{dk}
\end{equation}
where the inverse is defined via
$$\sum_k \int d\theta''\ K_{ik}(\theta,\theta'')
K^{-1}_{kj}(\theta'',\theta') = \delta_{ij} \delta(\theta-\theta').$$
Unfortunately, this reformulation,
while rather elegant, is also rather useless because
we cannot find a closed-form expression for $K^{-1}$ explicitly.
The problem in inverting (\ref{Kdef})
is that $K$ is not a function of the difference $\theta-\theta'$ but
of both the variables individually.
Observe from (\ref{dk}) that, as expected from general
considerations,
fluctuations of densities per unit length
are of order $O(1/L)$.

\subsection{Current noise in a Luttinger liquid}

We now use these general results to compute the current noise in
a particular system, the Luttinger liquid.
We first find the noise using conformal field theory, and then
rederive it using the quasiparticle picture.
This leads to a result which will be crucial to our analysis
in the next section including an impurity.
We find that
even though the current operator creates multi-particle states,
only one-particle states contribute to the
DC fluctuations.

The bulk Hamiltonian of the Luttinger model
can be written in terms of the charge currents \cite{Lutt}
\begin{equation}
H_0={ \pi\over g} \int_{0}^l dx
\left[j_L^2+j_R^2\right].
\label{hamil}
\end{equation}
The overall constant $g$ follows from the normalization of
the current; in the fractional quantum Hall problem this is
fixed by the charge of the electron and we have $g=\nu$ when
$1/\nu$ is an odd integer \cite{Wen}.
Since in this system the left and right movers are decoupled,
we here treat only the right movers.
We study the model at non-zero temperature $T$,
which corresponds to imaginary times on a
circle of circumference $1/T$.

Without an impurity, the charge and current
fluctuations in a Luttinger liquid (or any other
conformal field theory) can be found very straightforwardly
from conformal field theory. \cite{Wennoise}
The left- and right-moving current operators $j_L(x+t)$ and
$j_R(x-t)$ are individually conserved.
To find the noise, we need to integrate the two-point function
of the currents.
Conformal invariance fixes it to be
\begin{equation}
\langle j_R(x,t) j_R(0) \rangle = {\pi^2 T^2 g \over [\sinh (2\pi^2 T
(x-t))]^2}
\label{confcurrent}
\end{equation}
and likewise for the left movers (The strange normalization
in (\ref{confcurrent}) arises because of our choice $h=1$ instead of
the
more common $\hbar=1$. Also, we are ignoring subtleties coming from
the chiral anomaly
which vanish in the infinite-volume limit. \cite{pk})
The DC current fluctuations are then given simply by
\begin{eqnarray}
\nonumber
\overline{(\Delta I)^2} &\equiv&
\lim_{\omega\to 0}\int dt\ e^{i\omega t}{1\over 2}\langle \{j_R(x,t),
j_R(0,0)\}\rangle+\ (L\to R)\\
&=&  2gT,
\label{jn}
\end{eqnarray}
where we added the contribution of left movers.
This is
the well-known Johnson-Nykqvist formula (for spinless electrons),
because the conductance
$G$ of this Luttinger system is equal to $g$ (see \cite{Wen,KF} for
example).

We now rederive this result for the current
noise from the quasiparticles.
We have discussed the quasiparticles for this system at
length in \cite{FLS,FLSbig}.
The values of $g$ where the scattering is diagonal are
$g={1/t}$, $t$ an integer. We therefore restrict
to these values, although we expect
that the final results will extend
straightforwardly to general coupling.
There are $t$ different types of quasiparticles;
the kink (labeled by $+$), the antikink ($-$), and
the breathers ($1\dots t-2$).
Since there is no scale in the theory,
these right-moving
particles are massless, and have the dispersion relation
$e_i=p_i= m_i e^{\theta}$,
where $m_{+}=m_{-}=m/2$ and $m_j=m \sin[\pi j/(2(t-1))]$.
The overall scale $m$ is arbitrary and cancels out of any physical
quantities.
The explicit form of the kernels $\Phi_{jk}$ in the Luttinger
liquid is discussed for example in \cite{FLSbig},
where we also give a simpler form of the thermodynamic
Bethe ansatz equations (\ref{tbaeqs}).

The general fluctuations depend on space and time. To compute them
using the quasiparticles, one would like to be able to recover
expressions such as (\ref{confcurrent}) based on the action of
current
operators on multiparticle states.  Unfortunately the theory,
although
it appears ``almost free'' from the Bethe-ansatz point of view, is
truly interacting, and physical observables act on the quasiparticle
basis in a very complicated fashion.  Typically, operators like the
current or the energy density have non-vanishing matrix elements
(so-called form factors \cite{Smir}) between all pairs of states with
identical number of kinks minus number of antikinks; for instance,
$j_R(x,t)$ acting on the vacuum can create an arbitrary number of
breathers and pairs of kinks and antikinks. (Operators that are
simple
in the quasiparticle basis, such as the density of kinks, are highly
non-local in terms of the original physical current $j_R$, and
presumably of no physical interest.) This makes the computation of
the
space- and time-dependent fluctuations rather difficult, although not
impossible. \cite{LSS}

While local operators are quite complicated, their integral
over space can become much simpler.  A typical example is the
stress-energy tensor, whose integral gives rise to the
energy and momentum operators, both of which act diagonally on the
multiparticle states. Another example is the
$U(1)$ charge, $Q=\int j_R dx$. With the convention
that the original physical electrons have Coulomb charge $1$, the
breathers have no charge, but the kink
has charge $q=+1$ and the antikink has charge $q=-1$.
In our quasiparticle basis, $Q$ acts
diagonally, and is a sum of one particle
operators:
$Q|\theta_1,\ldots\theta_n\rangle_{q_1\dots q_n}=\left(\sum
q_i\right)
|\theta_1,\ldots\theta_n\rangle_{q_1,\ldots,
q_n}=\left( L\int (P_+-P_-)d\theta\right)
|\theta_1,\ldots\theta_n\rangle_{q_1,\ldots,q_n}$. This property
holds because the sine-Gordon Hamiltonian (and thus its massless
limit) and the charge operator commute, so they can
therefore be simultaneously
diagonalized. It can also be painstakingly checked using general
formulas
for  matrix elements of the local current operator.\cite{Smir}

This makes the computation of the DC noise
much easier than the AC noise. The point is
 that DC properties
are obtained by integrating over all times, and since we deal with
chiral objects (objects which are functions of either $x-t$
or $x+t$), this is the same as integrating over space.
Thus DC current fluctuations are related to
charge fluctuations: for instance,
as immediately seen from (\ref{jn}), one has
$\overline{(\Delta I)^2}={1\over L}\overline{(\Delta Q)^2}$ (we set
the Fermi velocity equal to unity).

Charge fluctuations can straightforwardly
be computed using the formalism of the previous section.
For a given configuration characterized by densities $P_i$,
the total charge of the right movers is
\begin{equation}
Q=L \int (P_+-P_-)d\theta.
\label{Qdef}
\end{equation}
The effect of a finite DC voltage to add to the free energy a term
$-VQ/2$. This corresponds to chemical potentials of
$\mu_+=-\mu_-=V/(2T)$ and $\mu_j=0$ for $j=1\dots t-2$.  Charge
fluctuations can now be obtained by taking derivatives with respect
to
$V$, the rapidity-independent chemical potential.  Using (\ref{Qdef})
with (\ref{onept}) and (\ref{dd}) gives
\begin{eqnarray}
\nonumber
&&\overline{Q} (V)  =
-{2L}{\partial\overline{ F}\over\partial V}\\
&&\overline{(\Delta Q)^2} = -{4L T}{\partial^2
\overline{F}\over\partial V^2}
= 2T{\partial \overline{ Q}\over\partial V}
\label{qsquared}
\end{eqnarray}
We can find $\overline{Q}(V)$ directly from the TBA.  By comparing
the
derivative of (\ref{tbaeqs}) with (\ref{bethe}), one finds that when
$\mu$ is independent of $\theta$
\begin{equation}
\overline{n}_i(\theta,V) =
T{\partial\overline{\epsilon}_i(\theta,V)\over \partial\theta}.
\label{ntoe}
\end{equation}
Using the definitions of $n_i$ and $f_i$
gives
$$
\int d\theta\ \overline{P}_{\pm}(\theta)
=T\int d\theta\ {\partial\overline{\epsilon}_{\pm}
(\theta,V)\over \partial\theta}
{1\over 1 + e^{\overline{\epsilon}_{\pm}(\theta,V) -\mu_{\pm}}}
$$%
The integrand is a total derivative, and from (\ref{Qdef}) we see
that
$$\overline{Q}(V)=LT\left[\ln
{1+e^{-\overline{\epsilon}_- +(V/2T)}\over
1+e^{-\overline{\epsilon}_+ -(V/2T)}}
\right]^{\overline{\epsilon}(\infty)}_{\overline{\epsilon}(-\infty)}.
$$
We see from (\ref{tbaeqs}) that
$\overline{\epsilon}(\infty)=\infty$, and it is straightforward
to show that \cite{FLSbig}
$$
\exp\left[\overline{\epsilon}_\pm(-\infty)\right]
={\sinh[(t-1)V/(2Tt)]\over\sinh [V/(2Tt)]}.
$$
This gives $\overline{Q}(V)= VL/t$ and using (\ref{qsquared}) gives
\begin{equation}
\overline{(\Delta Q)^2} ={2LT} g
\label{stateii}
\end{equation}
for any $V$. Using $\overline{(\Delta I)^2}=
{1\over L}\overline{(\Delta Q)^2}$
reproduces the desired result (\ref{jn}).

For completeness,
we can use this computation to give a quick proof of the
diagonal action of $Q$. For any multiparticle state which
we denote by  $|\cal{M}\rangle$, characterized  densities $P_i$,
one has
\begin{eqnarray}
\nonumber
&&\langle{\cal M}|\int dx\, j_R(x,t)\int dx\, j_R(x,t') |{\cal
M}\rangle
= \quad X\\
&&
\nonumber
\qquad +\ L^2\int d\theta d\theta'(P_+(\theta)-P_-(\theta))
(P_+(\theta')-P_-(\theta')).
\end{eqnarray}
Here $X$ represents the other contributions to the two-point
function:
\begin{equation}
X=\sum_{M'}\left|\langle{\cal M}|\int dx j_R(x)|{\cal
M}'\rangle\right|^2 e^{i(E_M-E_M')t},
\label{junkx}
\end{equation}
and $E_M$ is the energy of the multiparticle state $|{\cal
M}\rangle$. It is convenient to define
\begin{equation}
D(\theta,\theta')\equiv \overline{\Delta(P_+-P_-)(\theta)
\Delta(P_+-P_-)(\theta')}
\label{Ddef}
\end{equation}
We then have
\begin{eqnarray}
\nonumber
\overline{(\Delta I)^2}&=&L\int \int D(\theta,\theta')
d\theta d\theta' +{1\over L} \overline{X}\\
\nonumber
&=&{1\over L}\overline{(\Delta Q)^2}+ {1\over L} \overline{X}\\
&=& 2g{T} + {1\over L} \overline{X},
\label{claim}
\end{eqnarray}
where the second equality follows from the definition of $Q$ in
(\ref{Qdef}), the last equality follows from (\ref{stateii}), and
$\overline{X}$ has the same expression as (\ref{junkx}) but with a sum
on intermediate states restricted to $E_M=E_M'$.  However, the
conformal field theory result (\ref{jn}) says that $\overline{(\Delta
I)^2}=2gT$, so $\overline{X}=0$.  Since each contribution to
$\overline{X}$ is a square modulus, and therefore positive, this means
that each of these contributions must vanish individually.  Therefore,
the complicated matrix elements of the local current all disappear
after spatial integration, and the charge operator, which is all we
need to compute DC current fluctuations, acts diagonally on
multiparticle states, and is a simple sum of one particle operators.
This fact will prove crucial in the next section, where we include an
impurity.

We also note that if one were to neglect the non-diagonal
density fluctuations, one does not obtain the Johnson-Nyquist
formula even in this simple situation with no impurity. This can
easily be checked, because if we were to treat
the particles as free but with a non-Fermi distribution function,
then  $K_{ij}^{-1}(\theta,\theta')
=-f_i \delta_{ij} \delta(\theta-\theta')$. Using this to find
$D(\theta,\theta')$ yields 
\begin{equation}
\overline{\Delta I^2}_{\hbox{approx}}
= 2T{\sinh(gV/2T)\cosh((1-g)V/2T)\over \sinh(V/2T)}.
\label{wrong}
\end{equation}
Except at $g=1/2$ (where the Luttinger liquid is equivalent
to a free fermion), this 
contradicts not only the Johnson-Nyquist formula but also
the simple physical result that when there is no impurity,
the left movers and right movers are not coupled, and the noise should
not depend on the voltage.

\section{The effect of an impurity on the noise}

In this section we add a single impurity to the Luttinger liquid.
This couples the left and right movers and allows backscattering.
An impurity at $x=0$ corresponds to adding a term \cite{Wen,KF}
$$H_B = \lambda \cos[\phi_L(0)-\phi_R(0)]$$
to the Hamiltonian; we have written the theory in terms of
a boson so that $j_L =-\partial_x\phi_L(x,t)/2\pi$ and
$j_R =\partial_x\phi_R(x,t)/2\pi$.
Basically, the idea is that
we express the noise in terms of properties of the system far from
the impurity combined with the $S$ matrix elements for tunneling
through the impurity. The former are given in the preceding section,
while the $S$ matrix elements are known.\cite{GZ}
The analysis is similar to one given for free electrons
\cite{Butt}; this is possible because of the simple action of $Q$
on multiparticle states explained
the previous section.

As previously discussed \cite{FLS,FLSbig},
the transport properties
of a Luttinger liquid with a  backscattering
impurity can be mapped
to a Luttinger system of right movers alone.
The backscattered charge $Q_L-Q_R$ due to
left-right tunneling in the original problem is proportional
to the charge $Q$ in the new problem. It is not conserved
because of the impurity scattering.

To start, we discuss the one-point function of the current
in the presence of the impurity, quickly rederiving
earlier results.\cite{FLS,FLSbig}
There are two types of asymptotic states; the in states
injected by the reservoir from the left,
and the out states, after scattering
through the impurity.
The out state
must have the same momentum and energy because the integrability
allows only elastic scattering, so a kink can scatter only into
a kink or antikink of the same rapidity.
The $S$ matrix element for a
particle
of rapidity $\theta$ and type $i$
to scatter off the impurity into a particle of type $j$ is denoted
by $S_{ij}(\theta)$.
These are written out in \cite{GZ}; all we
will need for the noise are
the unitarity relation
\begin{equation}
\sum_{k} S_{ik}^\dagger S_{kj}=\delta_{ij}
\label{unitar}
\end{equation}
and the magnitude
\begin{equation}
{\cal T}(\theta)\equiv |S_{++}(\theta)|^2
={1\over 1+e^{2(g-1)(\theta-\theta_B)/g}}
\label{smat}
\end{equation}
where $\theta_B$ parametrizes the impurity strength
and is related to $\lambda$.\cite{FLSbig}
In particular, it
follows from the unitarity relation that $|S_{+-}|^2=1-{\cal T}$.

The effect of the impurity is that
the asymptotic states
$|\theta\rangle_+$
and $|\theta\rangle_-$
are not individual eigenstates of
the Hamiltonian any more.
The single-particle eigenstates are now
\begin{eqnarray}
\nonumber
||\theta\rangle_+&\equiv&{1\over
\sqrt{2}}\left\{|\theta\rangle ^{in}_++S_{++}(\theta)
|\theta\rangle ^{out}_++S_{+-}(\theta)|\theta\rangle
^{out}_-\right\}\\
||\theta\rangle_-&\equiv&{1\over
\sqrt{2}}\left\{|\theta\rangle ^{in}_-+S_{--}(\theta)
|\theta\rangle ^{out}_-+S_{-+}(\theta)|\theta\rangle
^{out}_+\right\}.
\label{neweigenstates}
\end{eqnarray}
The unitarity of the $S$ matrix
yields $_i\langle\theta||\ ||\theta\rangle_j=\delta_{ij}$.
Describing multiparticle
eigenstates is a little more complicated; for our purpose it
will be sufficient to think of them as tensor products
of the states (\ref{neweigenstates})\cite{LSS}.

Although $Q$ acts diagonally on the asymptotic states, it acts
in a more complicated fashion on the eigenstates of the problem
with  impurity: this time, $Q$ and the Hamiltonian
do not commute and cannot be simultaneously diagonalized.
First one has
\begin{eqnarray}
\nonumber
_+\langle \theta||Q||\theta\rangle _+ &=&
{1\over 2}\left(1+|S_{++}(\theta)|^2-|S_{+-}(\theta)|^2\right)\\
&=&{\cal T}(\theta)
\label{prettygood}
\end{eqnarray}
and
\begin{equation}
\
_-\langle \theta||Q
||\theta\rangle _-=-{\cal T}(\theta).
\label{prettygoodi}
\end{equation}
 The charge acts on multi-particle states
by multiplying by the appropriate
densities.
Since we will be
interested
in the leading contributions to the current, we can neglect the
effect of the boundary when quantizing the allowed rapidities.
Therefore the densities are exactly those discussed in the
previous section. Since the impurity problem is again mapped on
 purely right movers,
the relation between DC fluctuations of the current and fluctuations
of the charge holds as before. It
follows that
\begin{equation}
\overline{I}(V) =\int \left[\overline{P}_+ (\theta,V)
-\overline{P}_-(\theta,V)\right] {\cal T}(\theta) d\theta.
\label{key}
\end{equation}
This is the central result of \cite{FLS,FLSbig}, where it
was however derived quite differently.

The current noise follows in a similar fashion.
The central result of this paper is that
\begin{eqnarray}
\nonumber
\overline{(\Delta I)^2} &=&
L\int\int
D(\theta,\theta')
{\cal T}(\theta)
{\cal T}(\theta') d\theta d\theta' \\
&+&
\nonumber
\int \left[\overline{P}_+(\theta)(1-\overline{f}_-(\theta))+
\overline{P}_-(\theta)(1-\overline{f}_+(\theta))\right]\\
&&\qquad \times
{\cal T}(\theta)[1-{\cal T}(\theta)]d\theta,
\label{conj}
\end{eqnarray}
where we remind the reader that $D(\theta,\theta')$
is the density-density correlator defined in (\ref{Ddef}). Again,
fluctuations
are computed without the impurity, its effect being negligible at
leading order
in $L$.
The first term in (\ref{conj}) is the equivalent of (\ref{claim}),
with the
${\cal T}$ terms arising from (\ref{prettygood}) and
(\ref{prettygoodi}).
The second term  is an additional contribution
at coincident rapidities. It occurs because
$Q$ is not diagonal on the eigenstates.
Since $Q$ is diagonal on the asymptotic states, the only
additional contribution due to intermediate states
is proportional to
$$ _+\langle \theta||\, Q\, ||\theta\rangle _-\,
_-\langle \theta||\, Q\, ||\theta\rangle _+
={\cal T(\theta)}(1-{\cal T(\theta)}),$$
and another with $+\leftrightarrow -$.
This correction does not occur if
the multiparticle state has a double occupancy
at a particular rapidity  $\theta$ (a state of the form
$||\ldots,\theta,\theta\ldots\rangle_{\ldots,+,-,\ldots}$),
because then the intermediate state
$||\ldots,\theta,\theta\ldots\rangle_{\ldots,-,-,\ldots}$
has two particles in the same state,
which is not allowed by the Bethe equations. This gives rise to the
additional densities in the second
term of (\ref{conj}), whose equilibrium value is
$$\overline{P_+(\theta)(1-f_-(\theta))} +
\overline{P_-(\theta)(1-f_+(\theta))}.$$
Of course, using the usual
arguments (see the appendix)
the fluctuations in this term are suppressed by a
factor of $1/L$, so we can replace it with the product
of the individual equilibrium values, as in
(\ref{conj}). Since fluctuations of densities per unit length
are of order $O(1/L)$, both terms in (\ref{conj}) are of order
$O(1)$.

$Q$ acts diagonally on multiparticle
states as a sum of one particle operators. In the presence
of the impurity, $Q$ acts again as a sum of one-particle
operators, albeit non-diagonally. For each rapidity, there
are two possible states $||\theta\rangle_+$ and $||\theta\rangle_-$,
and in this subspace $Q$ acts as a two-by-two matrix. This matrix
can alternatively be represented by introducing fictitious
fermion creation and annihilation operators
as
\begin{eqnarray}
&&\left[a_+^\dagger(\theta)a_+(\theta)
-a_-^\dagger(\theta)a_-(\theta)\right]|S_{+-}(\theta)|^2\\
\nonumber
&&\qquad - a_+^\dagger(\theta)a_-
(\theta)S_{++}^*S_{+-}-a_-^\dagger(\theta)a_+(\theta)
S_{+-}^*S_{++}
\label{frombutt}
\end{eqnarray}
We can thus represent the DC current as an integral over rapidities
with (\ref{frombutt}) as the integrand. This formally
coincides with the free-electron calculation in \cite{Butt}. There
are however major  differences since populations
at different rapidities are correlated. For completeness,
let us briefly rederive
(\ref{conj}) in this approach.
Since we neglect the ($O(1/L)$)  effect of the impurity
on the densities, averages have to be computed  using  the asymptotic
states and the Bethe equations for the bulk system.  One then
checks that
for any such states, $\langle a_+^\dagger(\theta)a_-(\theta)\rangle
=0$
and
$\langle a_+^\dagger(\theta)a_+(\theta')
a_-(\theta')a_-^\dagger(\theta)\rangle =0$
unless $\theta=\theta'$. The first term of (\ref{conj}) follows right
away. For the
second term, we need also to observe that
$\langle
a_+^\dagger(\theta)a_+(\theta)a_-(\theta)a_-^\dagger(\theta)\rangle =
\langle a_+^\dagger(\theta)a_+(\theta)\rangle
\langle a_-(\theta)a_-^\dagger(\theta)\rangle +O(1/L)$,
and the $P_+(1-f_-)$ and $P_-(1-f_+)$ follow for the coincident
term.

We finally discuss how to compute (\ref{conj})
numerically. The second term is straightforward; the
densities follow from the TBA equations (\ref{tbaeqs})
and the relation (\ref{ntoe}).
There is a trick to make the numerics even easier;
as indicated in sect.\ 4.6 below, this piece can
be simply written as a derivative of the current.
The first term is much more
difficult. Extracting $D(\theta,\theta')$
from (\ref{dk}) is difficult
because it would require inverting
large matrices and taking their derivatives numerically.
Taking functional derivatives to determine $D(\theta,\theta')$
directly from (\ref{dd}) is also difficult.
However, if we choose the rapidity-dependent chemical potentials
to be
$$\mu_+(\theta)=-\mu_-(\theta)={V\over 2T}+x|S_{++}(\theta)|^2,$$
this defines a partition function $Z(x)$. Then
\begin{equation}
L\int\int D(\theta,\theta')
{\cal T}(\theta){\cal T}(\theta')
d\theta d\theta'=-{1\over T}
{\partial^2 \overline{F}\over\partial x^2}\Bigg|_{x=0}
\label{deriv}
\end{equation}
Using this method, we calculated the noise explicitly at $g=1/3$.
Some curves are given in fig.\ 1. A
more physical way to express the impurity coupling
is as $\theta_B\equiv \ln [T_B/T]$,
where $T_B$ is a crossover parameter analogous
to the Kondo temperature. (This definition
corresponds to setting the arbitrary scale $m=T$.)
The noise $\overline{(\Delta I)^2}$
is then a function of three variables $V,T$ and $T_B$.
Since $\overline{(\Delta I)^2}/T$ is dimensionless, it
is a function of two variables, say $V/T$ and $T_B/T$.
In fig.\ 1 each curve is a function of $T_B/T$ at
a fixed value of $V/T$. 
We notice that as $V/T$ is increased,
a peak develops, similarly to the conductance \cite{FLSbig}. However,
this peak appears at a lower value of $V/T$ than does the peak
in the conductance. 
\begin{figure}
\epsfxsize=3.5in
\epsffile{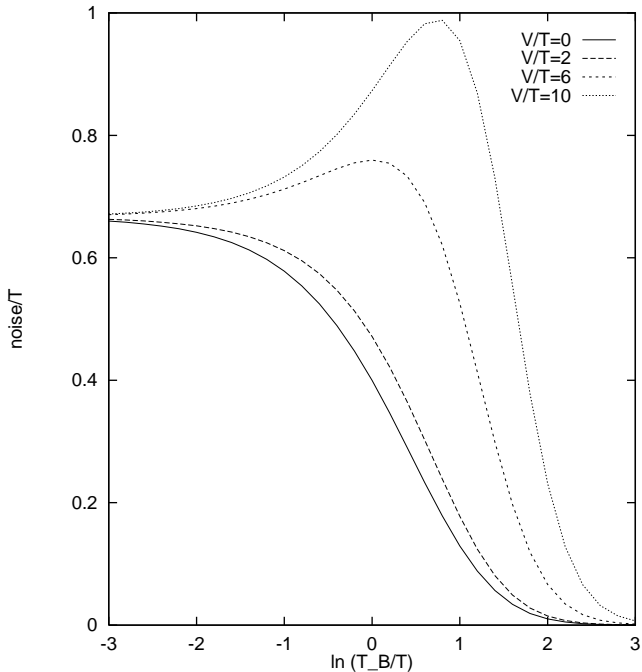}
\caption{The noise $\overline{(\Delta I)^2}/T$
as a function of impurity coupling for four different values of
$V/T$, where the coupling $g=1/3$.}
\end{figure}
\section{Various Limits}

We now consider  various limits
of (\ref{conj}) to check our expressions and
to investigate further physical aspects of the noise in the
interacting system.
A first trivial limit  is of course the case with no impurity,
where ${\cal T}=1$ and
the results of section 3 reduce to those of section 2.

\subsection{$T=0$}

At vanishing temperature, there are no fluctuations, so
$D(\theta,\theta')=0$; the noise is pure shot noise.
For positive voltage, $f_-=0$,
so the noise
reads simply
\begin{equation}
\overline{(\Delta I)^2} =\int
\overline{P}_+(\theta)\left[{\cal T(\theta)}-{\cal
T(\theta)}^2\right]d\theta,
\end{equation}
in agreement with \cite{FLSnoise}. There it is also shown that
a simple fluctuation-dissipation-type relation relates
the noise to the current. Unfortunately, we have not been
able to find a finite-temperature analog of this relation.

\subsection{$V=0$}

When the voltage $V=0$, we show that
the charge fluctuations $D(\theta,\theta')$ are diagonal and
find that the Johnson-Nyquist formula holds.
We first observe that due to the symmetry of the
kernels $\Phi_{ij}=\Phi_{ji}$,
the Bethe equations (\ref{bethe}) and the TBA equations
(\ref{tbaeqs}) yield
$$\overline{n}_+=\overline{n}_-\equiv \overline{n} \qquad\quad
\overline{\epsilon}_+
=\overline{\epsilon}_-\equiv\overline{\epsilon}$$
at any voltage.
Similarly, when $\mu_+(\theta)=-\mu_-(\theta) \equiv\mu
(\theta)$, we have
$$
\overline{P}_+[\mu(\theta)]=\overline{P}_-[-\mu(\theta)],
$$
from which it follows that
\begin{equation}
\overline{n}[\mu(\theta)]=\overline{n}[-\mu(\theta)]\qquad\quad
\overline{\epsilon}[\mu(\theta)]=\overline{\epsilon}[-\mu(\theta)].
\label{parity}
\end{equation}
We see from section 2 that $D(\theta,\theta')$ can be written
as the functional derivative
%
%
\begin{equation}
D(\theta,\theta')\big|_{V=0}=L {\delta
(\overline{P}_+(\theta)-\overline{P}_-(\theta))
\over\delta\mu(\theta')}
\Big|_{\mu(\theta)=0}.
\label{wayii}
\end{equation}
Because $\overline{P}_i=\overline{n}
\overline{f}_i$ and $\overline{n}$ and $\overline{\epsilon}$ are even
in $\mu$, the only non-vanishing term in the functional
derivative at $\mu=0$ arises from the
direct dependence
of the filling fractions (\ref{deff}) on $\mu$. Hence,
at vanishing voltage,
\begin{equation}
D(\theta,\theta')\big|_{V=0}
={2\over L}\ \overline{n}\ \overline{f}\
(1-\overline{f}\ )\delta(\theta-\theta').
\label{statei}
\end{equation}
Therefore, the
charge fluctuations are diagonal and free in the absence of
an applied  voltage. This fact also
follows easily from the symmetry of the TBA equations
under $+\leftrightarrow -$ when $V=0$, combined with (\ref{dk}).
One then finds
\begin{equation}
\overline{(\Delta I)^2} \big|_{V=0}=2\int \overline{n}\overline{f}
(1-\overline{f}){\cal T} d\theta
={2T}G,
\end{equation}
where we used the linear-response result for the
conductance $G$. \cite{FLS}
Thus
in limit $V=0$ we recover the Johnson-Nyquist formula.
This is an extremely non-trivial check on our result,
because the conductance here depends on the impurity coupling
and temperature.

\subsection{Free fermions}

When $g=1/2$, the Luttinger liquid is equivalent to free fermions.
The bulk $S$ matrices are unity, so the kernels $\Phi_{ij}=0$ and
the pseudoenergy $T\overline{\epsilon}_i(\theta)={m_i} e^{\theta}$.
Then $K_{ij}(\theta,\theta')=-(1/f_i)\delta_{ij}\delta(\theta-\theta')$
is easy to invert, yielding
$$
\overline{D(\theta,\theta')}=
{1\over L}
\left[\overline{P}_+
(1-\overline{f}_+)+\overline{P}_-(1-\overline{f}_-)
\right]\delta(\theta-\theta')
$$
and using $n_+=n_-\equiv n$, we have
\begin{eqnarray}
\nonumber
\overline{(\Delta I)^2} &=&\int  \overline{n}
\left\{\left[\overline{f}_+
(1-\overline{f}_-)+\overline{f}_-(1-\overline{f}_+)\right]{\cal
T}\right.
\\
\nonumber
&&\qquad -\left. (\overline{f}_+-\overline{f}_-)^2{\cal T}^2
\right\}d\theta
\end{eqnarray}
in agreement with the known results \cite{Land,Butt}.

This limit illustrates the importance of the fluctuations.
We note that if one were to use this formula for $g\ne 1/2$
as an approximation for the correct formula (\ref{conj}), one
obtains an answer larger than the correct one, even
if we use the appropriate interacting distribution functions
derived from the TBA.
For example,
as ${\cal T}\to 1$, one obtains (\ref{wrong}) instead
of the correct $2gT$.
The difference between the two is substantial for $V\ne 0$,
especially when $g$ is small.

\subsection{Strong-backscattering limit}

{}From (\ref{smat})
we see that for $\theta<<\theta_B$, the transmission amplitude
${\cal T}\to 0$. Thus
the limit $\theta_B\to\infty$ is the strong backscattering
limit, where the impurity splits the system in two.
In the expression (\ref{conj}) for the noise, in this limit
we can
neglect ${\cal T}^2$ compared to ${\cal T}$, except for very
energetic
particles with
$\theta\approx \theta_B$. However,
for $\theta\to\infty$ we have $\overline{\epsilon}\to\infty$
and $\overline{f}\to 0$, and
the contribution
of this region to the integral
is negligible. We can therefore approximate
(\ref{conj}) in this limit by neglecting the ${\cal T}^2$ terms
entirely:
$$\overline{(\Delta I)^2} \approx \int
\overline{n}\left[\overline{f}_+
(1-\overline{f}_-)+\overline{f}_-(1-\overline{f}_+)\right]
{\cal T} d\theta.
$$
By simple algebra one finds
$$
f_+(1-f_-)+f_+(1-f_-)=\coth\left({V\over 2T}\right)(f_+-f_-)
$$
Hence
$$\overline{(\Delta I)^2} \approx \coth\left({V\over 2T}\right)\int
\overline{n}\ (\overline{f}_+-\overline{f}_-){\cal T} =
\coth\left({V\over 2T}\right)\overline{I} .
$$
This is the noise coming from the tunneling
of uncorrelated electrons, a result expected on physical grounds
\cite{KFnoise}.

\subsection{Generalized-fluctuation dissipation results}

By the same arguments as in section 2,
the conductance $G=d\overline{I}/dV$ can be written as
\begin{equation}
G={L\over 2T}\int D(\theta,\theta') {\cal T}(\theta)
d\theta d\theta'.
\label{flscond}
\end{equation}
The
voltage difference $v$ between the two channels is given by
\cite{KFnoise}
$$v={1\over g}({I_{max}}-I),$$ where ${I_{max}}$
is current with no impurity.
Since the noise in $v$ is proportional
to the noise in the reflected (as opposed to transmitted)
current, it is therefore given by the same arguments of section 3
with the the transmission amplitude ${\cal T}$ replaced
with the reflection amplitude $1-{\cal T}$.
Therefore
\begin{eqnarray}
\nonumber
\overline{(\Delta v)^2} &=&
L\int\int
D(\theta,\theta')
(1-{\cal T}(\theta)
(1-{\cal T}(\theta')) d\theta d\theta' \\
&+&
\nonumber
\int \left[\overline{P}_+(\theta)(1-\overline{f}_-(\theta))+
\overline{P}_-(\theta)(1-\overline{f}_+(\theta))\right]\\
&&\qquad \times
{\cal T}(\theta)(1-{\cal T}(\theta))d\theta,
\label{vconj}
\end{eqnarray}
By simple
algebra, and using (\ref{flscond}),
we derive the following identity
\begin{equation}
\overline{(\Delta I)^2} =g^2\overline{(\Delta v)^2} +2T(2G-g).
\end{equation}
This is the zero-frequency limit of the general relation
between current and voltage fluctuations, equation (3) of
\cite{KFnoise}.

\subsection{Weak-backscattering limit}

The weak-backscattering limit is given by $\theta_B \to -\infty$.
This is not as easy to treat as the strong backscattering limit,
because
the fluctuations for very-low-energy particles are not negligible,
preventing
a naive expansion in $1-{\cal T}$.
We therefore have not succeeded in checking
the weak backscattering limit analytically.
Some steps can be accomplished however.
We can evaluate the behavior of the second
term in (\ref{conj}) using the identity
$$
{\cal T}(1-{\cal T})= {g\over
2(g-1)}{\partial\over\partial\theta_B}{\cal T}.
$$
Therefore
\begin{eqnarray}
\nonumber
&&\int \overline{n}\ \left[\overline{f}_+(1-\overline{f}_-)
+\overline{f}_+(1-\overline{f}_+)\right]{\cal T}(1-{\cal T})
d\theta\\
\nonumber
&&={g\over 2(g-1)}\coth ({V\over 2T}) \int
\overline{n}\ (\overline{f}_+-\overline{f}_-)
 {\partial\over\partial\theta_B}{\cal T}\\
&&={g\over 2(g-1)}\coth ({V\over 2T}){\partial
\overline{I}\over\partial\theta_B}.
\end{eqnarray}
In the weak backscattering limit, the current
can be expanded in a power series $\overline{I}
=\overline{I}_{max}+c(V,g,T)
e^{2(1-g)\theta_B}$ (where $\overline{I}_{max}=gV$), so
$${\partial
\overline{I}\over\partial\theta_B}=
2(1-g)(\overline{I}-\overline{I}_{max})
$$
and
\begin{eqnarray}
\nonumber
&&
\overline{(\Delta I)^2}\approx g \coth(V/(2T))
(\overline{I}_{max}-\overline{I})+2T(2G-g)\\
\nonumber
&&\qquad +\int\int D(\theta,\theta')
(1-{\cal T}(\theta) )(1-{\cal T}(\theta'))
d\theta d\theta'.
\end{eqnarray}
On the other hand, in this limit physical arguments say
that the noise comes from the uncorrelated
tunneling of Laughlin quasiparticles of charge $g$,
yielding \cite{KFnoise}
$$\overline{(\Delta I)^2} \approx
 g\coth(gV/2T) (\overline{I}_{max}-\overline{I})+2T(2G-g).$$
This will hold if and only if
\begin{eqnarray}
\nonumber
&&L
\int\int D(\theta,\theta')(1- {\cal T}(\theta))(1- {\cal T}
(\theta'))d\theta d\theta'\\
&&\qquad
\approx g\left[\coth({gV\over 2T})-
\coth\left({V\over 2T}\right)\right]
(\overline{I}_{max}-\overline{I}).
\label{weakb}
\end{eqnarray}
This can be checked explicitly in the case $g=1/2$, but in
general we need to resort to a numerical check. For $g=1/3$ we
computed
this integral by expressing it as the derivative of $Z(x)$, as
discussed at the end of section 3. We found $Z(x)$ by numerically
solving the TBA equations, and took the derivatives numerically.
We find that (\ref{weakb}) is
indeed satisfied for any voltage as one takes $\theta_B\to \infty$.

\section{Conclusion}

We have presented in this paper an exact computation of the
non-equilibrium noise
in a Luttinger liquid with a single impurity.
The approach is somewhat similar
to the one for free electrons
because the model is integrable. However, the quasiparticles
of this integrable model are not free, and the computation
requires as a key ingredient
the  equilibrium fluctuations of densities, which are not diagonal
and are quite complicated.

The computation of AC properties appears more difficult. In addition
to the correlations between densities at different
rapidities induced by the Bethe equations, another key ingredient to
take
into account is that local physical operators are infinite sums of
multiparticle
operators with complicated matrix elements (form factors). We hope
however that a computation similar to \cite{LSS} is possible.

\bigskip
\bigskip

We would like to thank A. Ludwig for previous collaborations
and interesting conversations.
This work was supported by the Packard Foundation, the
National Young Investigator program (NSF-PHY-9357207) and
the DOE (DE-FG03-84ER40168).

\appendix\section{Density
fluctuations from the thermodynamic Bethe ansatz}

The energy per unit length
is given simply by summing up the individual
energies $e_i(\theta)$ of the particles:
$$E=\int e_i(\theta) P_i(\theta) d\theta.$$
In order to compute the entropy, we utilize the fact
that solutions of the Bethe equations allow
only one quasiparticle per level,
so that the filling fractions obey $0\le f_i \le 1$.
The number of states in the interval between $\theta$ and
$\theta + d\theta$ is then
$$(n_i(\theta)d\theta)!\over
(P_i(\theta)d\theta)!([n_i(\theta)_-P_i(\theta)]d\theta)!.$$
Stirling's formula gives the entropy
$$S=
\sum_i\int d\theta \left[ n_i\ln n_i - P_i\ln P_i -
(n_i-P_i)\ln (n_i-P_i)\right].$$
A given configuration
is completely characterized by the knowledge of the densities
$\{P_i(\theta)\}$,
or by the knowledge of the pseudoenergies $\{\epsilon_i(\theta)\}$
defined in (\ref{deff}). These sets are individually complete
because the Bethe equation (\ref{bethe}) gives the $n_i$ in terms
of the $P_i$, and so knowing the $\epsilon_i$ determines the $P_i$
and vice versa.
Either
set can be used as convenient variables for the functional
integration. In what follows
we keep dependent variables in intermediate computations for
notational simplicity.

The various quantities are obtained by a functional integration with
the Boltzmann weight
$e^{-FL/T}$, where $F=E-T S$. As discussed
in sect.\ 2, the full action includes a chemical potential, and
it can be written as
\begin{eqnarray}
\nonumber
&&F=\sum_i\int (e_i-\mu_i)\ P_i\ d\theta-
T\sum_i
\int P_i\left[\ln(1+e^{\epsilon_i-\mu_i})\right.
\\
\nonumber
&&\qquad\qquad \left.
+e^{\epsilon_i-\mu_i}\ln
(1+e^{-\epsilon_i+\mu_i})\right]d\theta,
\label{action}
\end{eqnarray}
where we generalize the customary situation by allowing
rapidity-dependent chemical potentials $\mu_i(\theta)$.

The equilibrium values $\overline{\epsilon}_i$ or  $\overline{P}_i$
minimize the action (\ref{action}) subject to the constraint
of the Bethe equations (\ref{bethe}). The first-order variation
gives
\begin{eqnarray}
\nonumber
&&{\Delta F}=\sum_i \int e_i \Delta P_i-
\sum_i\int P_i
e^{\overline{\epsilon}_i-\mu_i}
\ln(1+e^{-\overline{\epsilon}_i+\mu_i})
\Delta\epsilon_i\\
\nonumber
&& -
T\sum_i\int
\left[\ln(1+e^{\overline{\epsilon}_i-\mu_i})+
e^{\overline{\epsilon}_i-\mu_i}\ln
(1+e^{-\overline{\epsilon}_i+\mu_i})\right]
\Delta P_i.
\end{eqnarray}
Recall that the $\Delta P_i$ are given in terms of $\Delta
\epsilon_i$
via the Bethe equations.
Requiring $\Delta F=0$ gives
the TBA equations (\ref{tbaeqs}) with the equilibrium free energy
(\ref{feq}).

To compute the fluctuations, we expand this action to second
order around the saddle point values (denoted with
a bar). Writing $\Delta^{(2)} P_i$ for the variation of
$P_i$ induced
by first and
second order variations in $\epsilon$ we get
\begin{eqnarray}
\nonumber
&&{\Delta^{(2)}F}/T =\sum_i \int e_i \Delta^{(2)} P_i/T\\
\nonumber
&-&
\sum_i\int P_i
e^{\overline{\epsilon}_i-\mu_i}
\ln(1+e^{-\overline{\epsilon}_i+\mu_i})
\Delta\epsilon_i\\
\nonumber
&-&
\sum_i\int
 \Delta^{(2)} P_i
\left[\ln(1+e^{\overline{\epsilon}_i-\mu_i})
+e^{\overline{\epsilon}_i-\mu_i}\ln
(1+e^{-\overline{\epsilon}_i+\mu_i})\right]
\\
\nonumber
&-&\sum_i\int P_i e^{\overline{\epsilon}_i-\mu_i} \left[
\ln(1+e^{-\overline{\epsilon}_i+\mu_i})
-{1\over
1+e^{\overline{\epsilon}_i-\mu_i}}\right]{(\Delta\epsilon_i)^2
\over 2}.
\end{eqnarray}
The definition of $\epsilon$ (\ref{deff}) relates the fluctuations
of $n,P$ and $\epsilon$:
$$e^{-\overline{\epsilon}_i+\mu_i}\Delta^{(2)}
(n_i-P_i)=\Delta^{(2)} P_i +\overline{P}_i\left[
\Delta\epsilon_i +{1\over
2}(\Delta\epsilon_i)^2
\right].$$
Also, from the Bethe equations (\ref{bethe}) we have
$$\Delta^{(2)} n_i=\sum_j\Phi_{ij}*\Delta^{(2)} P_j.$$
It follows that $\Delta^{(2)} P_i$ depends
only on the combination $\Delta \epsilon_j+{1\over
2}(\Delta\epsilon_j)^2$.
Because the first-order variation $\Delta F=0$,
the linear terms
in $\Delta\epsilon$ and many others
cancel out, leaving only
$${\Delta^{(2)} F}={1\over 2}\sum_i\int
{\overline{P}_i\over 1+e^{-
\overline{\epsilon}_i+\mu_i}}(\Delta\epsilon_i)^2.$$
(A similar result appears in the original paper by Yang and Yang
\cite{YY} on
bosons with repulsive delta interactions). Hence the fluctuations of
$\epsilon$ are diagonal:
\begin{equation}
\langle \Delta \epsilon_i(\theta)\Delta\epsilon_j(\theta')\rangle
={1+e^{-\overline{\epsilon}_i+\mu_i}\over
L \overline{P}_i}\delta_{ij}
\delta(\theta-\theta').
\label{epfluc}
\end{equation}
In terms of the filling fractions $f_i$
we can recast this as
$$\langle \overline{n}_i(\theta)\Delta f_i(\theta)
\overline{n}_j(\theta')\Delta f_j(\theta')\rangle ={1\over L}
\overline{n}_i\overline{f}_i(
1-\overline{f}_i)\delta_{ij}\delta(\theta-\theta').$$
This is the {\bf same} result as one would have for a free theory:
fluctuations
restricted to the filling fractions are free. The difference with a
free theory
is that we have correlated fluctuations
$\langle \Delta n_i\Delta n_j\rangle $ as
well as $\langle \Delta n_i\Delta f_j\rangle $.

The density-density fluctuation $D(\theta,\theta')$ can be
obtained from (\ref{epfluc}) by expressing $\Delta\epsilon_i$ in
terms of the $\Delta P_i$. The Bethe ansatz equations (\ref{bethe})
yield
$$(1+e^{\overline{\epsilon}_i-\mu_i})\Delta P_i +
e^{\overline{\epsilon}_i-\mu_i}\overline{P}_i
\Delta\epsilon_i=\sum_j\Phi_{ij}
*\Delta P_j,$$
which can be recast as
$$
e^{\overline{\epsilon}_i-\mu_i}\overline{P}_i\Delta\epsilon_i=
\sum_j \int K_{ij}(\theta,\alpha)\Delta P_j(\alpha)d\alpha,$$
where $K_{ij}$ is defined in (\ref{Kdef}).
Plugging this into (\ref{epfluc}) gives an explicit form
of $D(\theta,\theta')$ in terms of an integral of $K^{-1}$:
\begin{eqnarray}
\nonumber
\langle &&\Delta P_i(\theta)\Delta P_j(\theta')\rangle
\\
\nonumber
&& ={1\over L}\sum_k\int d\alpha\ K^{-1}_{ik}(\theta,\alpha)
K^{-1}_{jk}(\theta',\alpha)
e^{\overline{\epsilon}_k(\alpha)-\mu_k}
\overline{n}_k(\alpha).
\end{eqnarray}
This equation can be simplified.
First note that when $\mu$ is independent of $\alpha$,
(\ref{ntoe}) yields
$$e^{\overline{\epsilon}_k(\alpha)}
\overline{n}_k(\alpha)=T {\partial
e^{\overline{\epsilon}_k(\alpha)}\over
\partial \alpha}.$$
Then it follows from the definition of $K$ that
$$
{\partial e^{\overline{\epsilon}_k(\alpha)}\over \partial \alpha}
\delta(\alpha-\alpha') \delta_{km}= -\left({\partial\over\partial
\alpha}
+{\partial\over\partial \alpha'}\right)
K_{km}(\alpha,\alpha').$$
Using these two identities and the definition of the inverse
of $K$ yields
the relation (\ref{dk}) relating the density fluctuations to
$K^{-1}$.

\end{document}